\documentclass[11pt]{iopart}
\RequirePackage{filecontents}
\usepackage{inputenc}
\usepackage{textcomp}
\usepackage[numbers,sort&compress]{natbib}

\def\url#1{}
\setcitestyle{square}
\setcitestyle{comma}
\usepackage{hyperref}

\usepackage[a4paper]{geometry}
\usepackage[inline]{enumitem}

\usepackage{graphicx}
\usepackage{subcaption}

\begin{document}

\pagestyle{plain}

\graphicspath{ {./figures/} }

\title{Machine learning methods for the detection of polar lows in satellite mosaics: major issues and their solutions.}
\author{Mikhail Krinitskiy$^1$, Polina Verezemskaya$^1$, Svyatoslav Elizarov$^2$, Sergey Gulev$^1$}
\ead{krinitsky@sail.msk.ru}
\address{$^1$ Shirshov Institute of Oceanology, Russian Academy of Sciences, Moscow, Russia}
\address{$^2$ Alterra.ai, Moscow, Russia}
\vspace{10pt}

\begin{indented}
\item[]December 2019
\end{indented}

\begin{abstract}
Polar mesocyclones (PMCs) and their intense subclass polar lows (PLs) are relatively small atmospheric vortices that form mostly over the ocean in high latitudes. PLs can strongly influence deep ocean water formation since they are associated with strong surface winds and heat fluxes. Detection and tracking of PLs are crucial for understanding the climatological dynamics of PLs and for the analysis of their impacts on other components of the climatic system. At the same time, visual tracking of PLs is a highly time-consuming procedure that requires expert knowledge and extensive examination of source data.

There are known procedures involving deep convolutional neural networks (DCNNs) for the detection of large-scale atmospheric phenomena in reanalysis data that demonstrate a high quality of detection. However, one cannot apply these procedures to satellite data directly since, unlike reanalyses, satellite products register all the scales of atmospheric vortices. It is also known that DCNNs were originally designed to be scale-invariant. This leads to the problem of filtering of the scale of detected phenomena. There are other problems to be solved, such as low signal-to-noise ratio of satellite data, and an unbalanced number of negative (without PLs) and positive (where a PL is presented) classes in a satellite dataset.

In our study, we propose a deep learning approach for the detection of PLs and PMCs in remote sensing data, which addresses class imbalance and scale filtering problems. We also outline potential solutions for other problems, along with promising improvements to the presented approach.
\end{abstract}

\section{Introduction}
\sectionmark{Introduction}
\label{intro-setion}
Polar mesocyclones (PMCs) are high-latitude atmospheric vortices that form mostly over the ocean. PMCs are relatively small and short-living: their sizes vary from 200 to 1000 km, and lifetime typically does not exceed 36 hours \cite{verezemskaya_southern_2017}. The intense subclass of PMCs, namely polar lows (PLs) characterizes by strong winds and high surface fluxes \cite{rasmussen_polar_2003}. Despite their small sizes, PLs can cause rough seas and may deliver dangerous weather conditions for engineering infrastructure, mostly due to their explosive formation and unpredictable behavior. Moreover, PMCs and PLs were shown recently to influence ocean convection, and deep water formation significantly \cite{marshall_open-ocean_1999, condron_modeling_2008, condron_impact_2013}. Therefore, reliable identification and tracking of PLs and PMCs are vital for their quantification as well as for assessing their impact on the ocean circulation of various scales.

The identification of PLs in atmospheric reanalyses is limited due to their sizes compared to effective grid resolutions. The lack of representation of PMCs in modern reanalyses was shown recently \cite{verezemskaya_southern_2017, condron_impact_2013, laffineur_polar_2014, michel_polar_2017, bromwich_comparison_2016}. As for reanalyses of high enough resolution, the number of studies demonstrating PLs identification is still insufficient (e.g., \cite{gavrikov-JAMC-RAS-NAAD}). In some studies, automated mid-latitude cyclone tracking methods are applied to operational analysis data with some adaptations to PL identification and tracking \cite{zappa_can_2014, pezza_southern_2016, xia_comparison_2012}. However, the reported climatological characteristics such as the number of PLs, their sizes, their lifetime vary significantly in these studies.

Due to the lack of representation of PLs in reanalyses and operational analyses, in our study, we focus on the detection of PLs in satellite imagery. Particularly, we use satellite imagery of cloudiness for the identification of cloudiness patterns associated with PLs. There are a number of studies exploiting the approach of manual identification and tracking of PLs \cite{verezemskaya_southern_2017, harold_mesocyclone_1999-1, harold_mesocyclone_1999, carrasco_mesoscale_1997, turner_summer-season_1994, carleton_interpretation_1995}, resulting in a comprehensive evaluation of characteristics of PLs in various regions. However, the procedure of manual identification of PLs requires expert knowledge and is enormously time-consuming. Thus, most of these studies are regional and cover relatively short time periods. This approach is unacceptable in case one would like to build a global long enough climatology of PLs. In this paper, we present further development of the study presented in \cite{krinitskiy_convolutional_2018} which main goal is the automated method for the identification and tracking of PLs in satellite data, which can reliably mimic a human expert. This method will tackle the issue of high costs of PL identification.

Machine learning (ML) and deep learning (DL), in particular, were shown recently to be quite effective in problems related to pattern recognition, visual object detection, and other computer vision tasks \cite{lin_feature_2016, simonyan_very_2014, badrinarayanan_segnet:_2015, he_deep_2016, krizhevsky_imagenet_2012}. Some studies demonstrated successful applications of DL methods in problems of the recognition of extreme weather events in reanalyses \cite{liu_application_2016, muszynski_topological_2019, rupe_towards_2019}. However, most of the existing studies are focused on the detection of synoptic-scale phenomena such as tropical cyclones or atmospheric rivers. DL methods were also shown to be applicable in the problem of the detection of mesoscale and sub-mesoscale oceanic eddies in SAR satellite imagery \cite{huang_deepeddy:_2017}. However, as of our best knowledge, there are no applications of DL methods for the identification of PLs in satellite imagery.

There are several reasons for this lack of studies:
\begin{enumerate*}[label=(\roman*)]
    \item A low signal-to-noise ratio of satellite imagery limits the efficiency of most computer vision algorithms.
    \item Satellite imagery preserves a full range of scales of atmospheric vortices, including synoptic cyclones, MCs, and sub-mesoscale vortices, so the lack of the ability of scale filtering may suppress the quality of an algorithm of PL detection.
    \item One of the most challenging problems of DL applications in supervised problems in climate sciences is the lack of labeled datasets. While there are many well-established labeled datasets and benchmarks for computer vision tasks \cite{deng_imagenet:_2009, krizhevsky2009learning, lecun_gradient-based_1998, cordts_cityscapes_2016, lin_microsoft_2015, everingham_pascal_2015}, climate sciences are characterized by a huge amount of unlabeled remote sensing data, reanalysis products, and model outputs. There are just a few labeled datasets of PLs that may be used as supervision for DL models \cite{verezemskaya_southern_2017, noer2010dates, noer_climatological_2011}. In addition to the limited number of these catalogues, their sizes are small compared to the number of labels typically needed for state of the art DL methods.
    \item Even in the case of the dataset created with a consistent methodology applied to a homogeneous remote sensing dataset \cite{verezemskaya_southern_2017}, one has to deal with imbalanced labels meaning that the area occupied by MCs ("positive" labels of imagery pixels) is low compared to the rest area ("negative" labels of imagery pixels).
\end{enumerate*}

Most of the mentioned issues do not impact the quality of the detection of large-scale extreme weather events in reanalysis data (e.g., \cite{liu_application_2016}). However, each of them has to be addressed in case of the identification of PLs in satellite imagery.

In this study, we present the novel method for the identification of PLs in satellite imagery using a specific adaptation of an existing deep learning approach addressing the issues of 
\begin{enumerate*}[label=(\alph*)]
 \item labels imbalance;
 \item scale filtering;
 \item small labeled dataset
\end{enumerate*}.
We also outline further advances in our study, which may improve the quality of the detection.

The rest of the paper is organized as follows: in section \ref{data-section}, we describe the source data and the database of PLs we use; in section \ref{method-section}, we overview the preprocessing of the source data and our method for the detection of PLs that we developed based on deep convolutional neural networks (DCNN); in section \ref{results-section}, we present the results of the application of our methodology. In section \ref{conclusions-section}, we summarize the paper with the conclusions and provide an outline of our further study.

\section{Data} \label{data-section}
In this study, we used the database of PMCs in the Southern Ocean (SOMC, \url{http://sail.ocean.ru/antarctica/}) described in detail in \cite{verezemskaya_southern_2017}. The database contains 1735 MC tracks resulting in 9252 records of MC size, position, and cloudiness pattern (according to the given in \cite{carleton_interpretation_1995}). SOMC database covers the summer season of 2004 (June to September). The trajectories of the database were built visually by an expert applying the methodology described in \cite{carleton_interpretation_1995}. The source data for this database were consecutive three-hourly IR ($10.3-11.3$ microns) and WV ($\sim6.7$ microns) satellite mosaics provided by the Antarctic Meteorological Research Center (AMRC) - Antarctic Satellite Composite Imagery (AMRC~ASCI). The spatial resolution of the mosaics is 5 km, and temporal resolution is 3 hours for the year 2004. AMRC~ASCI mosaics cover the area to the South of $\sim40^{\circ}$S (see complete description in \cite{kohrs_global_2014}).

\begin{figure}
     \centering
     \begin{subfigure}[b]{0.32\textwidth}
         \centering
         \includegraphics[width=\linewidth]{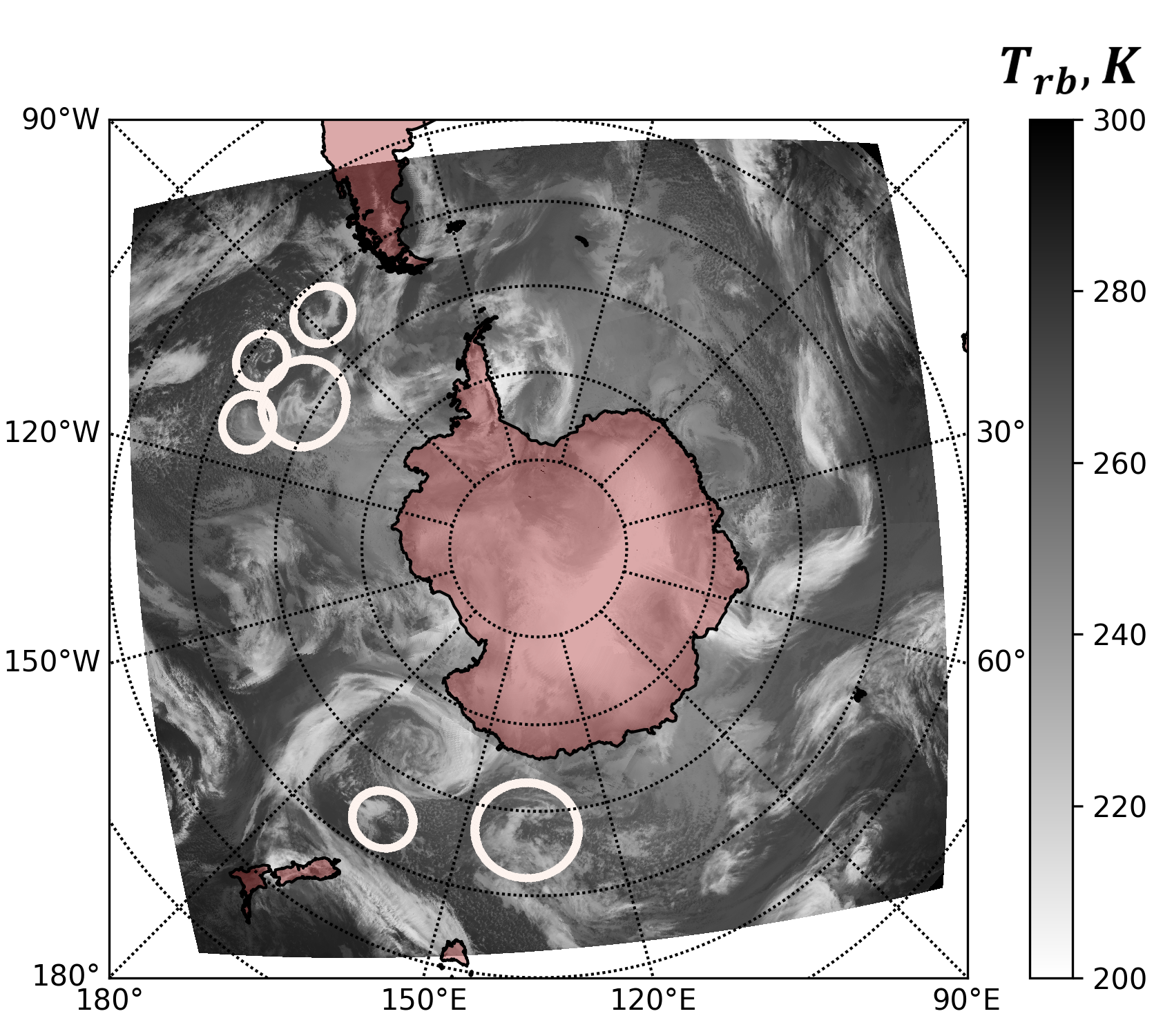}
         \caption{}
         \label{fig_1:subfigure_a}
     \end{subfigure}
     \hfill
     \begin{subfigure}[b]{0.32\textwidth}
         \centering
         \includegraphics[width=\linewidth]{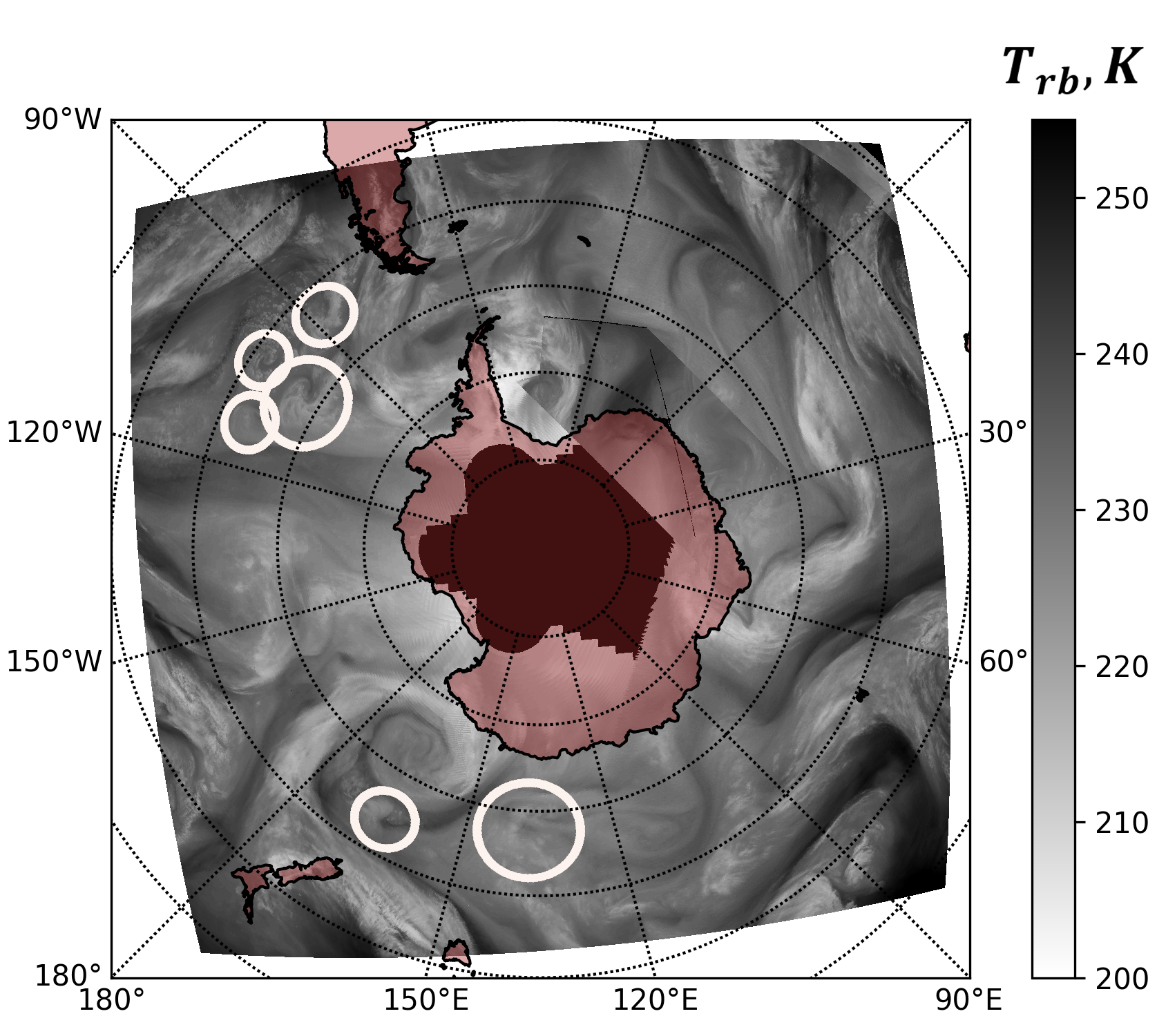}
         \caption{}
         \label{fig_1:subfigure_b}
     \end{subfigure}
     \hfill
     \begin{subfigure}[b]{0.32\textwidth}
         \centering
         \includegraphics[width=\linewidth]{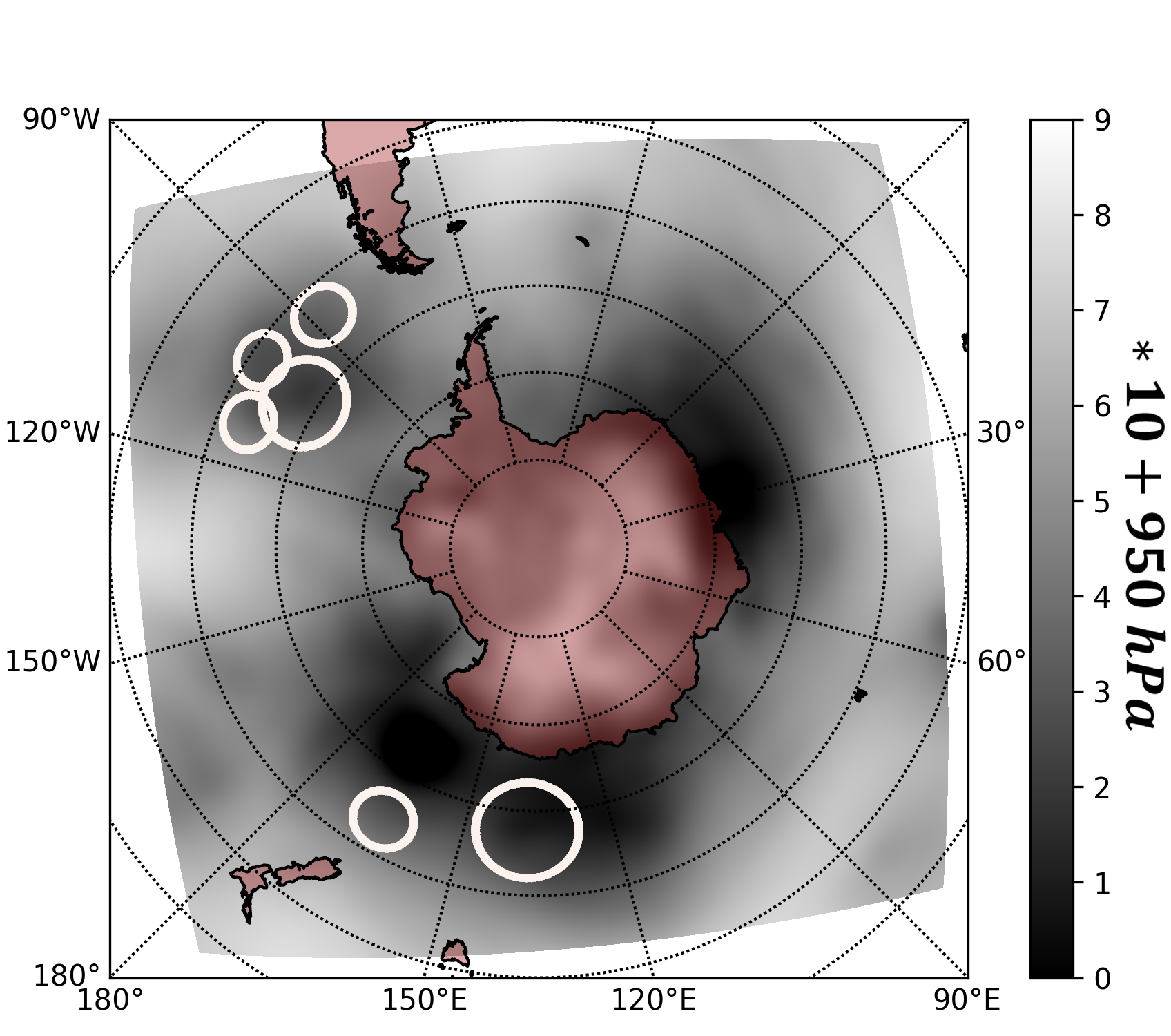}
         \caption{}
         \label{fig_1:subfigure_c}
     \end{subfigure}
\caption{Source data at 2004-06-01 00:00 (UTC) for the detection of PLs: (a) IR mosaic, (b) WV mosaic, and (c) SLP snapshot linearly downscaled to AMRC~ASCI grid. The white circles mark PLs identified by an expert in this snapshot.}
\label{fig_1}
\end{figure}

For the development of the detection algorithm, we processed the same AMRC~ASCI satellite imagery of IR and WV channels. We also included spatially and temporally collocated sea level pressure (SLP) field of ERA-Interim reanalysis \cite{dee_era-interim_2011}, which we interpolated to the AMRC~ASCI grid. SLP here essentially represents large-scale circulation patterns e.g. synoptic cyclones, which one needs to filter out at the moment of PLs detection. In fig. \ref{fig_1}, one particular snapshot of source data is presented.

\section{Method} \label{method-section}

The main problem of this study is to infer PLs positions and sizes, i.e., bounding boxes of PLs (bboxes hereafter) for each time moment meaning in each 3-hourly source data snapshot (see examples in fig. \ref{fig_3} shown by red squared bboxes).

\subsection{Data preprocessing} \label{method-section-data-preprocessing}
First, we normalize the source data according to eq. \ref{eq_norm}:

\begin{equation} \label{eq_norm}
    x_{norm} = \frac{x-x_{min}}{x_{max} - x_{min}},
\end{equation}
where $x$ denotes IR, WV, or SLP fields; $x_{min}$ and $x_{max}$ are shown in the tab. \ref{tab_norm_quantiles}. With this normalization procedure, 99\% of the values of normalized source data are in the range [0, 1]. All the values outside this range were masked and thus did not contribute to the loss function of our DCNN model. Also, source satellite data comes with a mask that marks the pixels where satellite data is missing or not usable.

\begin{table}[h!]
\centering
\begin{tabular}{c c c}
 \hline
 Feature & $x_{min}$ & $x_{max}$ \\ [0.5ex] 
 \hline\hline
 IR & 200 K & 300 K \\ 
 WV & 200 K & 255 K \\
 SLP & 950 hPa & 1040 hPa \\ [1ex] 
 \hline
\end{tabular}
\caption{Normalizing coefficients for source data}
\label{tab_norm_quantiles}
\end{table}

For the computational purpose, we cut sub-regions of source data of size 1024x1024 pixels for the training procedure of our DCNN. The size of one grid cell of AMRC~ASCI mosaics is $\sim5$ km, so the size of these cut regions is $\sim5120$ km, which is sufficient for fitting any known PL. In fig. \ref{fig_3}, examples of these sub-regions are shown. We perform this sub-sampling procedure with the condition that each of these sub-regions contains at least one PL bbox, and all the PL bboxes were present in these sub-regions completely. In fig. \ref{fig_3}, these ground truth labels (PL bboxes) introduced by a human expert are marked by red bboxes.

We also apply data augmentation, which was shown to be helpful for DL models to generalize better \cite{dosovitskiy_discriminative_2016, krizhevsky_imagenet_2012, zeiler_visualizing_2014, howard_improvements_2013}. In our study, we applied multiple augmentation techniques, including Spatial Transform Networks \cite{jaderberg_spatial_2016} and Deep Diffeomorphic Transformer Networks \cite{skafte_detlefsen_deep_2018, freifeld_transformations_2017}. We also employed a set of weak affine transformations using the python library "imgaug" \cite{imgaug}. The augmentation was applied only to the train subset of data. As a result, its amount increased significantly. In tab. \ref{tab_datasets}, the number of augmented examples is shown for the train set, as well as the numbers of non-augmented sub-region examples and corresponding ground truth PL labels for validation and test sets.

\begin{table}[h!]
\centering
\begin{tabular}{c c c} 
 \hline
 Data subset & No. of sub-regions & No. of PL labels \\ [0.5ex] 
 \hline
 train (augmented) & 1009728 & 2466674 \\ 
 validation & 12800 & 35252 \\
 test & 9984 & 23482 \\ [1ex] 
 \hline
\end{tabular}
\caption{The number of cut sub-region examples and corresponding ground truth PL labels in subsets of source data.}
\label{tab_datasets}
\end{table}

\subsection{RetinaNet for the identification of PLs} \label{method-section-retinanet}
In our study, we use the adaptation of the particular artificial neural network called RetinaNet presented in \cite{lin_focal_2017}. This network is designed to perform the task of object detection, which is precisely the case of PL identification.

The issue of unbalanced datasets is well known in ML and DL problems. There are several ways to address this issue \cite{buda_systematic_2018}. In the problem of the detection of PLs in satellite imagery, it is barely possible to employ data-level methods (oversampling/undersampling). Thus, the only option is classifier-level methods. In \cite{lin_focal_2017}, cost-sensitive learning is employed \cite{buda_systematic_2018}. In particular, the novel loss function was introduced, namely focal loss (FL hereafter, see eq. \ref{eq_focal_loss} for the $\alpha$-balanced variant of FL).
\begin{equation} \label{eq_focal_loss}
    FL({\hat{p}}) = -\alpha(1-\hat{p})^{\gamma}*\log{(\hat{p})},
\end{equation}
where $\hat{p}$ is an output of the network modeling a Bernoulli-distributed variable that is commonly interpreted as an estimate of the probability of an object (enclosed with a detected bbox) to have class "1", i.e., to be a PL in our study; $\alpha$ is a balancing parameter; $\gamma$ is a focusing parameter of this loss function. FL was designed in replacement of the commonly used cross-entropy loss to tackle the issue of highly imbalanced labels, which is precisely the case in the problem of PL detection in satellite imagery. In this study, we adjusted the value of $\gamma=3.5$, which differs from the one proposed in \cite{lin_focal_2017}. This value enforces the focus of our DCNN on positive labels.

The architecture of the model RetinaNet was originally presented in \cite{lin_focal_2017}. Apart from the FL, a few novel approaches were implemented in RetinaNet. Among others, it exploits transfer learning (TL) \cite{kolesnikov_large_2019} with a backbone sub-network anticipatorily trained on a large dataset like ImageNet \cite{deng_imagenet:_2009}. In DL methods, TL addresses the lack of labeled data. In our study, ResNet-152 \cite{he_deep_2016} was employed as a backbone network. The authors of RetinaNet empathize though that it achieves top results in object detection tasks not based on innovations of the network design, but due to the novel loss function.

Following the established practice in visual object detection problems, we assess the performance of PLs detection by mAP metric (\cite{everingham_pascal_2015}; higher is better; its values are in the range $[0, 1]$). We also evaluated our adapted RetinaNet by mean intersection-over-union ($IoU$, see eq. 1 in \cite{everingham_pascal_2015}) over all the detected PLs.

Following the approaches described in RetinaNet paper \cite{lin_focal_2017}, we designed our DCNN with the abovementioned adaptations regarding FL parameters and structural particularities. The constructed DCNN based on RetinaNet was trained on the augmented train subset. Quality assessment was performed on the test subset. For each of the label proposals, the DCNN we constructed performs two tasks:
\begin{enumerate*}[label=(\roman*)]
 \item classification meaning estimating the probability $\hat{p}$ of the label proposal to enclose a PL, and
 \item regression meaning adjusting the position and the size of the label proposal
\end{enumerate*}. Then we filter these label proposals with the thresholding value $p_{th}$. In our study, $p_{th}$ is a hyperparameter. We optimized mean $IoU$ w.r.t. $p_{th}$ using validation subset of data. We achieved the optimal value $p_{th}=0.34$.

\section{Results and discussion} \label{results-section}
\begin{figure}
     \centering
     \begin{subfigure}[b]{0.32\textwidth}
         \centering
         \includegraphics[width=0.95\linewidth]{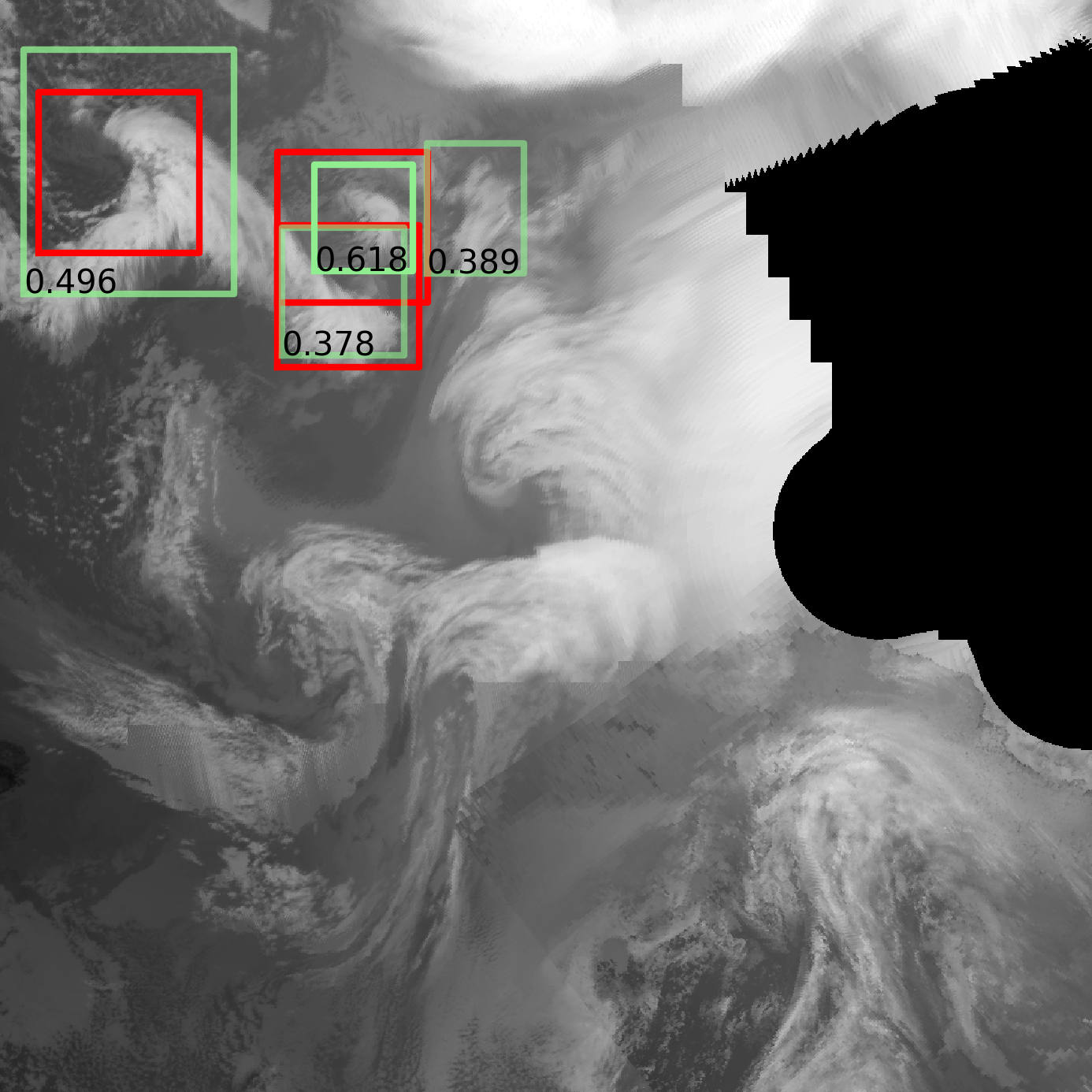}
         \caption{}
         \label{fig_3:subfigure_a}
     \end{subfigure}
     \hfill
     \begin{subfigure}[b]{0.32\textwidth}
         \centering
         \includegraphics[width=0.95\linewidth]{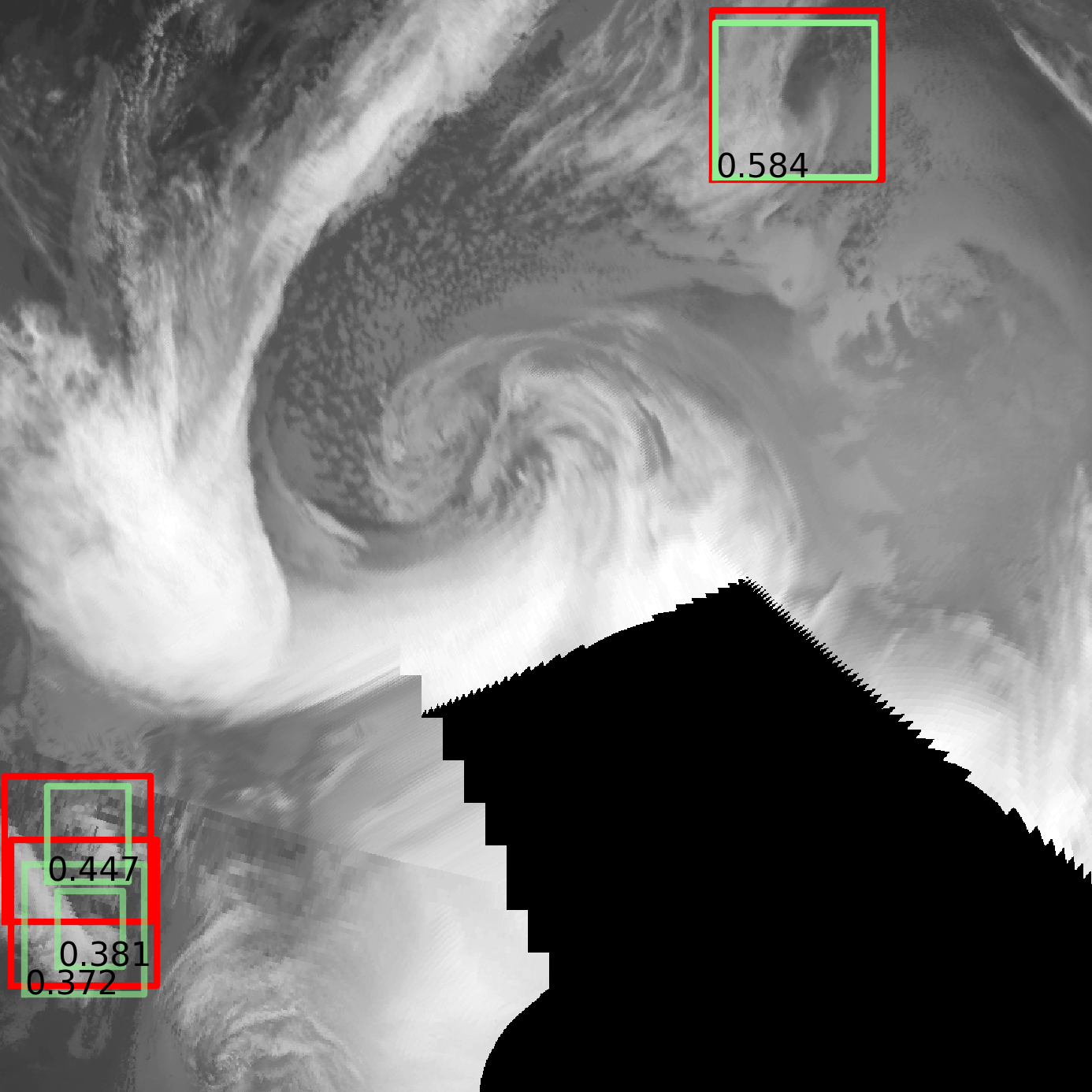}
         \caption{}
         \label{fig_3:subfigure_c}
     \end{subfigure}
     \hfill
     \begin{subfigure}[b]{0.32\textwidth}
         \centering
         \includegraphics[width=0.95\linewidth]{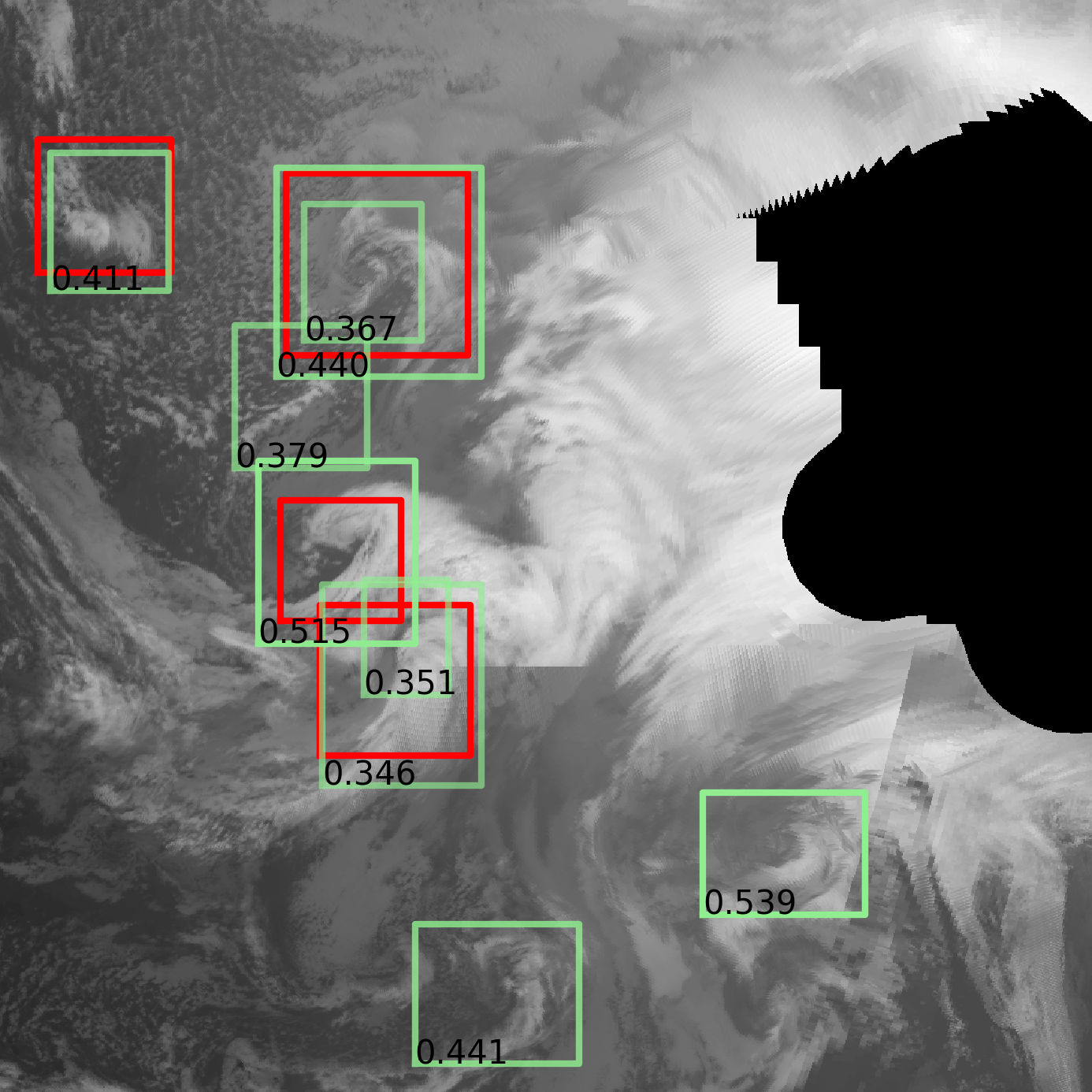}
         \caption{}
         \label{fig_3:subfigure_d}
     \end{subfigure}
     
\caption{Results of the proposed detection model. Background shown with grayscale color map represents IR normalized source data. Black areas represent the mask of missing data. Ground truth labels are marked with red rectangles; detected PLs are marked with green bounding boxes with the probability estimates $\hat{p}$ indicated by text labels.}
\label{fig_3}
\end{figure}

In fig. \ref{fig_3}, some examples of the application of the adapted RetinaNet are presented. As one can see, qualitatively, our DCNN performs the PL detection task in good accordance with ground truth. For example, in fig. \ref{fig_3:subfigure_c} one can see an almost perfect match of detected PLs with ground truth. In fig. \ref{fig_3:subfigure_d}, all the PLs were identified; however, some false alarms are presented. There are also a few PLs detected twice in the presented results.

With the $p_{th}$ optimized as mentioned in section \ref{method-section-retinanet}, we assessed the quality of PLs detection in terms of mean $IoU$ and $mAP$: $IoU_{mean} = 0.27; mAP = 0.36$. One may argue that despite the good qualitative correspondence of identified PLs with expert-defined labels, formal metrics are low compared to state of the art values (e.g., $0.869$ in \cite{singh_sniper:_2018}). However, there is a fundamental difference between a typical computer vision-based object detection task and the task of PL detection. In a common visual object detection, there are no concepts of the lifetime of objects, of the formation of objects or dissipation of objects. Thus, the comparison of $mAP$ metric values with the best ones achieved with state of the art DL approaches seems impractical.

\begin{figure}
     \centering
     \begin{subfigure}[b]{0.32\textwidth}
         \centering
         \includegraphics[width=0.95\linewidth]{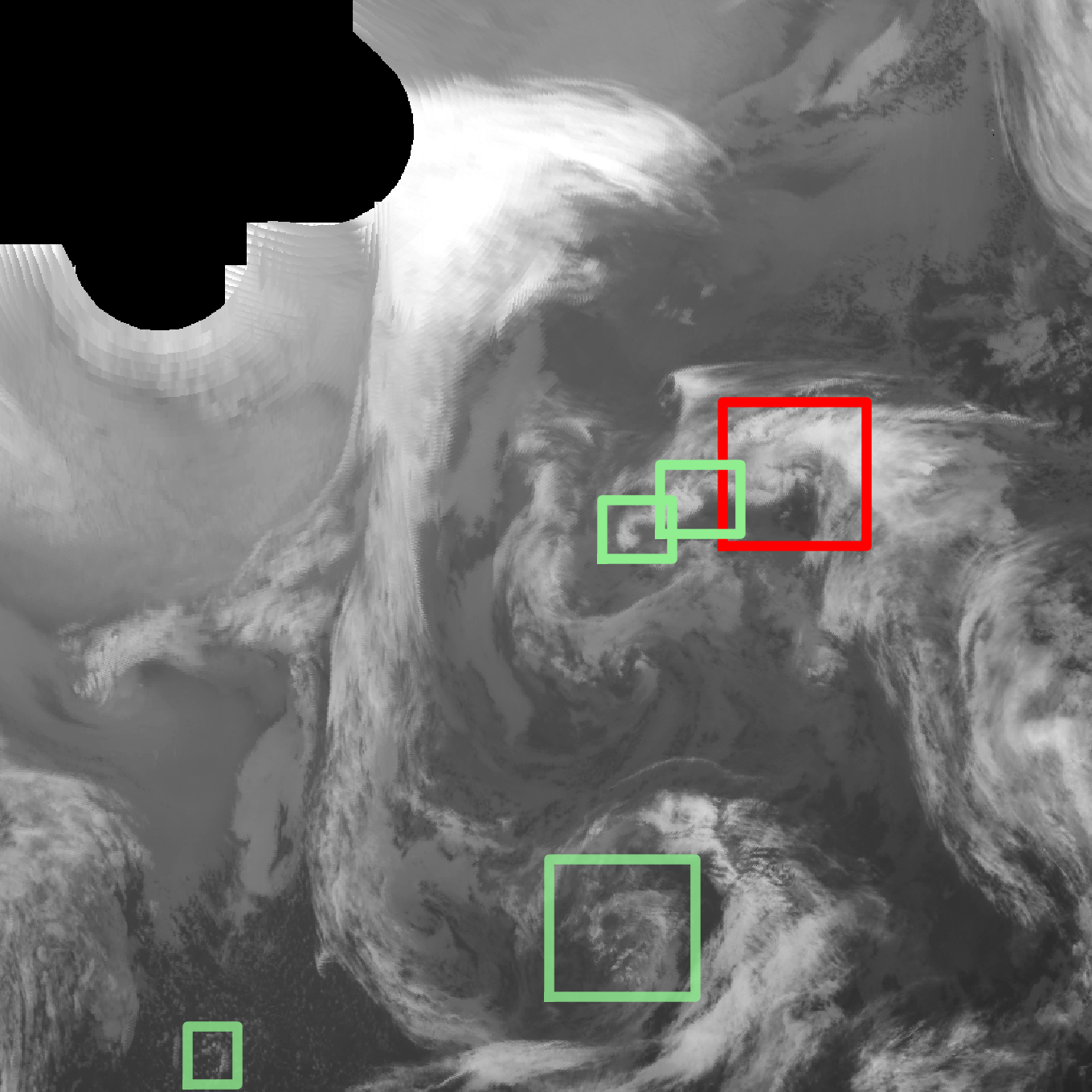}
         \caption{timestep t}
         \label{fig_4:subfigure_a}
     \end{subfigure}
     \hfill
     \begin{subfigure}[b]{0.32\textwidth}
         \centering
         \includegraphics[width=0.95\linewidth]{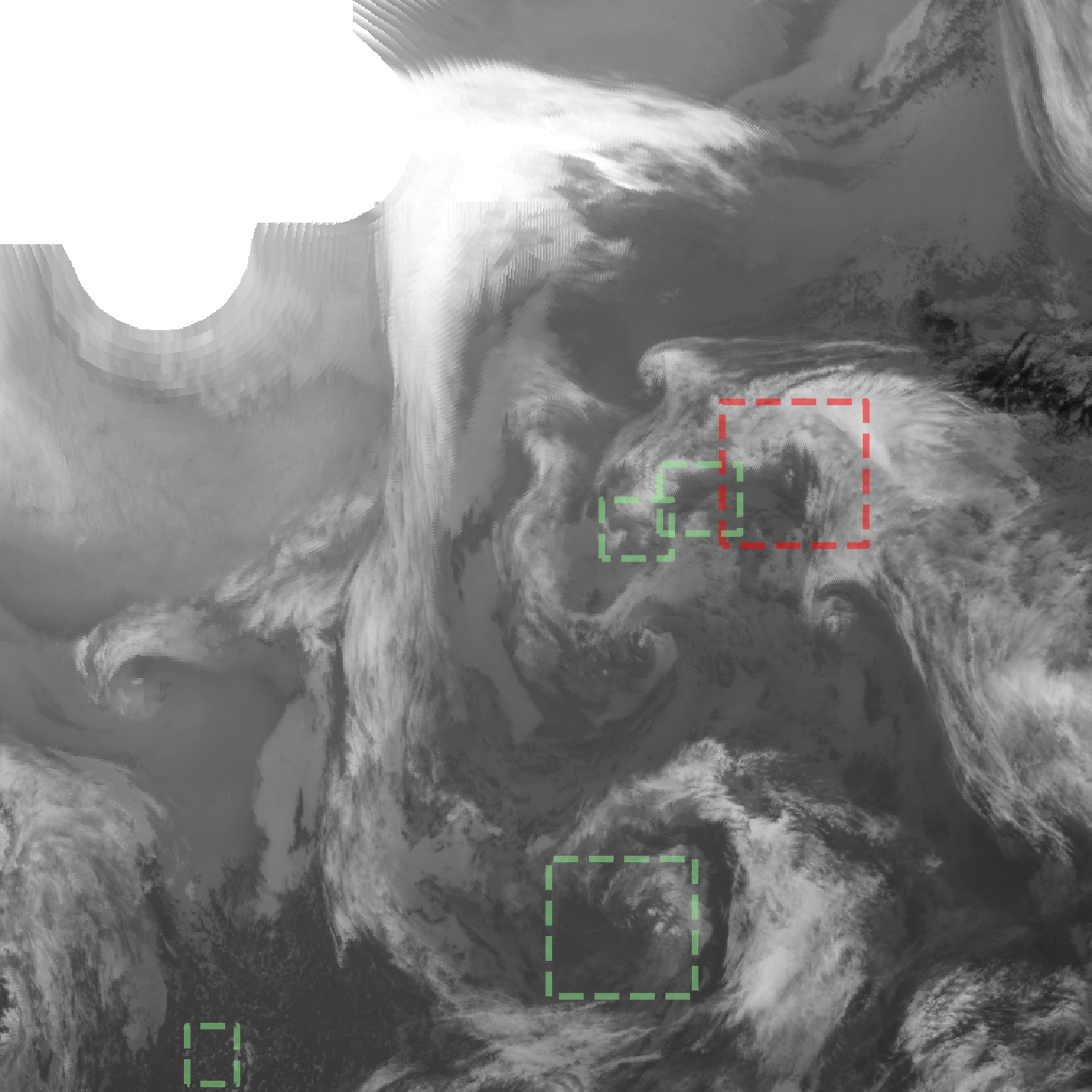}
         \caption{timestep (t+3h)}
         \label{fig_4:subfigure_b}
     \end{subfigure}
     \hfill
     \begin{subfigure}[b]{0.32\textwidth}
         \centering
         \includegraphics[width=0.95\linewidth]{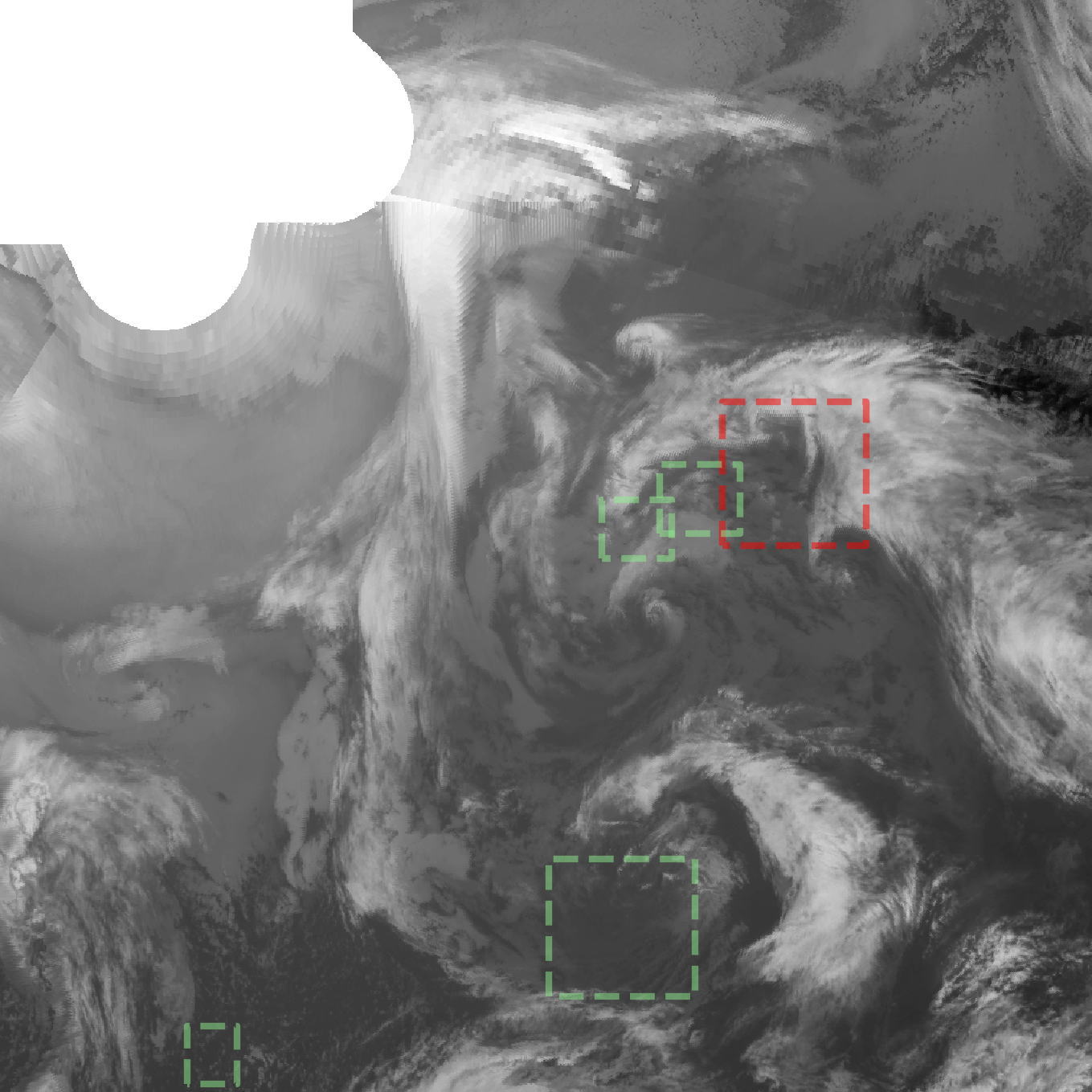}
         \caption{timestep (t+6h)}
         \label{fig_4:subfigure_c}
     \end{subfigure}
     
\caption{An example of false PL detections: (a) detected and true labels at a time step $t$; (b) the same labels at the same positions 3 hours after the time step $t$; (c) the same labels at the same positions 6 hours after the time step $t$.}
\label{fig_4}
\end{figure}

One of the main issues of PL detection problem compared to common object detection is the need for conditioning of PL identification on its temporal evolution. For example, consider fig. \ref{fig_4} with three consequent snapshots of source IR data. In fig. \ref{fig_4:subfigure_a}, the detected PLs (green bboxes) and expert-identified PLs (red bboxes) are presented; the same bboxes in the same positions are shown in figures \ref{fig_4:subfigure_b} and \ref{fig_4:subfigure_c} for a reference. According to the methodology \cite{carleton_interpretation_1995}, some of the detected cloudiness patterns may be interpreted as PLs. However, these PL candidates dissipate quickly, so there are almost no signs of these patterns in 3 and 6 hours. The lack of the proposed DCNN of consideration of temporal evolution results in falsely detected PLs (false alarms). Currently, we consider this issue as a major factor limiting the performance of our DCNN.

There are additional factors limiting the performance of the proposed method:
\begin{enumerate}[label=(\alph*)]
    \item in some cases of missing source data (shown with black regions in figures \ref{fig_3} and \ref{fig_4}), an expert may assume linear propagation of a PL through these regions, which results in an additional PL label inside the masked region. Currently, there is no way to implement or mimic this kind of decision in our method;
    \item the amount of labeled data in the PL detection problem is extremely insufficient for the application of DL approaches to this problem. The abovementioned data augmentation and transfer learning do improve the performance of the DCNN. However, additional human expert labeling strongly required.
\end{enumerate}

\section{Conclusions and outlook} \label{conclusions-section}
In this paper, a novel method for the identification of PLs in satellite mosaics is presented. This method is based on a DCNN named RetinaNet, which was proposed for the visual object detection task. As of our best knowledge, it is the first application of deep convolutional neural networks to the problem of the detection of PLs and PMCs in satellite mosaics data. In this study, RetinaNet and FL were adapted to the specifics of the problem of PL detection. In particular, focusing parameter $\gamma$ of the FL was set to a higher value in order to make the DCNN more focused on positive labels that denote the bboxes where PLs are presented. With this approach, we address the issue of class imbalance. The source data for the identification of PLs were satellite mosaics of AMRC~ASCI. We also included spatially and temporally collocated SLP fields from ERA-Interim reanalysis in order to address the issue of scale filtering since SLP essentially encodes large scale atmospheric circulation.

There are several approaches we intend to apply further. The most promising modification of the proposed model is a new neural network architecture that takes the temporal evolution of multiscale atmospheric circulation into account. There are additional modifications of the method, which may improve the quality of the detection. One of them is the regularization of the model, which constrains the geometric parameters of label bboxes. One more potential improvement implies the development of the SOMC database. The promising approach here is to use assisted labeling. Since there is already a model presented in this study that detects PLs and PMCs with some noise, it is promising to use it for generating noisy PL proposals. For a human expert, binary classification ("it is a PL"/"it is not a PL") of the proposed labels is much faster compared to an extensive examination of the whole source data example. This approach makes it possible to address the major deep learning issue of insufficient labeled datasets.

\section{Acknowledgements} \label{acknowledgements-section}
This work was undertaken with financial support by  the Russian Ministry of Science and Higher Education (agreement \textnumero 05.616.21.0112), project ID RFMEFI61619X0112)

\def\bibfont{\small}
\bibliography{references}

\end{document}